Review article

Lingling Huang, Shuang Zhang and Thomas Zentgraf*

# Metasurface holography: from fundamentals to applications

**Abstract:** Holography has emerged as a vital approach to fully engineer the wavefronts of light since its invention dating back to the last century. However, the typically large pixel size, small field of view and limited space-bandwidth impose limitations in the on-demand high-performance applications, especially for three-dimensional displays and large-capacity data storage. Meanwhile, metasurfaces have shown great potential in controlling the propagation of light through the well-tailored scattering behavior of the constituent ultrathin planar elements with a high spatial resolution, making them suitable for holographic beam-shaping elements. Here, we review recent developments in the field of metasurface holography, from the classification of metasurfaces to the design strategies for both free-space and surface waves. By employing the concepts of holographic multiplexing, multiple information channels, such as wavelength, polarization state, spatial position and nonlinear frequency conversion, can be employed using metasurfaces. Meanwhile, the switchable metasurface holography by the integration of functional materials stimulates a gradual transition from passive to active elements. Importantly, the holography principle has become a universal and simple approach to solving inverse engineering problems for electromagnetic waves, thus allowing various related techniques to be achieved.

*Corresponding author: Thomas Zentgraf,* Department of Physics, University of Paderborn, Warburger Straße 100, Paderborn 33098, Germany, e-mail: thomas.zentgraf@uni-paderborn.de. http://orcid.org/0000-0002-8662-1101
**Lingling Huang:** School of Optics and Photonics, Beijing Institute of Technology, Beijing 100081, China; and Department of Physics, University of Paderborn, Warburger Straße 100, Paderborn 33098, Germany
**Shuang Zhang:** School of Physics and Astronomy, University of Birmingham, Birmingham B15 2TT, UK

# 1 Introduction to metasurface holography

Holography, first invented in 1948 by Denis Gabor [1], is one of the most promising imaging techniques in wave phenomena and enables recording and reconstruction of the full wave information of a certain target or object wave. The hologram itself is not an image, rather, it consists of seemingly random patterns of spatially varying intensity or phase. In 1961, following the pioneering work of Gabor, Leith and Upatnieks applied the principle of holography to free-space optical beams with the advent of the laser [2]. The next important milestone was the invention of the computer-generated hologram (CGH) by Brown and Lohman in 1966 [3]. By using numerical computation to calculate the phase information of the wave at the hologram interface, they were able to simplify the recording process through programming. This breakthrough opened up many new possibilities, such as the creation of holograms of virtual objects that do not exist in reality, or the realization of dynamic holograms using computer-controlled spatial light modulators (SLM) [4]. In 1972, the concept of holography was extended by Cowan from free-space waves to surface waves [5]. Since then, holography-based techniques have been used to achieve three-dimensional (3D) displays, data storage, metrology, biological image processing and electron tomography interference, among others [6]. Holographic principles have been applied to overcome various challenges where other approaches have failed. However, there are still several remaining challenges for traditional holography technologies due to the bulky macro-scale interference-based generation methods, which have imposed fundamental physical limits for realizing the reconstruction, such as narrow bandwidth, small field-of-view (FOV), multiple diffraction orders, twin images and so on. These challenges mainly arise from the limitations of the CGH algorithm, the relatively large pixel sizes and the limited space-bandwidth product [7].

Nowadays, the enormous progress in nanofabrication techniques may revolutionize the method of achieving optical holography. The subwavelength tailored optical elements (gratings, etc.) and, more generally, metamaterials and metasurfaces are opening new frontiers for holographic and optical devices, such as beam splitters and image processing. Metamaterials are artificial structures engineered to show properties that cannot be found in natural materials. Even though metamaterials research started with the quest for negative-index and artificial magnetism, the field has grown into a much broader discipline encompassing various optical functionalities, in particular for the arbitrary control of light waves at the nanoscale [8, 9]. Metasurfaces – as a two-dimensional (2D) version of metamaterials – can provide powerful control over the flow of light through judiciously engineered planar nanostructures. Such a planar approach can serve as a promising route for practical applications. Metasurfaces usually feature a spatially varying optical response (e.g. scattering amplitude, phase and polarization) for molding optical wavefronts into almost any arbitrary profiles. Metasurfaces can also be integrated with functional materials to accomplish active control and greatly enhanced nonlinear response [10–13]. Meanwhile, metasurfaces have shown the high potential of being easily integrated into multifunctional on-chip optoelectronic systems in both optical and radio frequency ranges [14, 15]. The emergence of the so-called flat optics with metasurfaces has yielded groundbreaking phenomena in electromagnetics and photonics by overcoming the limitations of conventional optics [10–13]. A wide range of applications based on plasmonic or dielectric metasurfaces have been proposed and demonstrated in wavefront engineering [16, 17], information processing [18, 19] and spin-controlled photonics [20–22].

One of the cutting-edge nanotechnologies combines holography with nanodevices [23, 24]. Such metasurface holography can be performed by mapping the configuration precisely to the position and the local scattering properties of nanoscale optical resonators patterned at the interface. Compared with conventional holograms, metasurface holograms have three major advantages [25]. First, metasurfaces can provide unprecedented spatial resolution, low noise and high precision of the reconstructed images, given that both phase and amplitude information of the wavefront can be recorded in ultrathin holograms. Second, the significantly reduced subwavelength pixel size contributes to an improved holographic image compared to traditional holograms due to the elimination of undesired diffraction orders. Third, they can provide large space-bandwidth products due to their ability to achieve large area fabrication. The holography principle has emerged as a viable tool in designing novel plasmonic interfaces to excite either free space beams or surface waves with desired field distributions.

In the current work, we review recent advancements in metasurfaces holography. We provide an overview of key holographic concepts and procedures for processing metasurface holograms. We also introduce the basic physical mechanisms and classifications of metasurfaces for a better understanding of the working principles. Both free-space waves and surface plasmon polariton waves can be controlled in an arbitrary manner by using the holographic principle. We introduce several holographic multiplexing strategies, as well as some pioneering approaches for achieving active holographic displays. Finally, we identify some promising areas for future research and indicate likely avenues for future metasurface device development. Tables 1 and 2 provide an overview of the state-of-art development of metasurface holography.

**Table 1:** Categories of the metasurface holograms.

| Category | Mechanism | Representative works | Operation wavelength/efficiency |
|---|---|---|---|
| Phase-only holograms | Effective medium theory | Multilayer I-shaped metamaterials [26] | 10.6 μm; 1.14% |
|  | Pancharatnam-Berry phase principle | Metasurface hologram with 80% efficiency [27] | Visible to near infrared; 80% |
|  | Huygens' metasurfaces | Huygens' metasurfaces composed of nanodisks [28] | 1600 nm; 90% |
| Amplitude-only holograms | Binary' amplitude modulation | Random photon sieve with nanoholes [29] | Visible range; 47% |
|  | Resonance effect | V-shaped antennas by tailoring the geometry's parameters [30] | 676 nm; 10% |
| Complex amplitude holograms | Resonance effect together with the Pancharatnam-Berry phase principle | C-shaped antenna array by tailoring the arc lengths and orientation angles [31] | 0.3–1 THz; 6.4% |
|  | Huygens' metasurface | Nanodisks with spatially variant lattice period [32] | 1477 nm: 40% |

Table 2: Multi-functionalities of the metasurface holography.

| Functionality | Mechanism | Representative works |
|---|---|---|
| Holographic multiplexing | Wavelength multiplexing | Multicolor dielectric metasurface consisting of the multiplexed Si nanoblocks [33] |
| | Polarization multiplexing | Simultaneous polarization and phase modulation by using the tailored elliptical posts [34] |
| | Hybrid multiplexing algorithm | Multiple recording channels, such as the position, polarization and angle [35] |
| Surface wave holography | Holographic interference pattern (by etching equal phase contours) | Four-fold polarization-controlled surface wave holographic multiplexing [36] |
| | PB phase principle | Spin-selective control of SPP holography by using the metasurfaces composed of nanoapertures [37] |
| Nonlinear holography | Nonlinear PB phase for multiple channel display | Spin- and wavelength-dependent nonlinear holographic multiplexing with SRR arrays [38] |
| | Resonance effect | Double-layer V-shaped antennas for third-harmonic generations of holography [39] |
| Active holography | 2D materials | Active metasurface holograms composed of functional graphene oxide materials [40] |
| | Phase change materials | Active metasurface holograms integrated with GST [41] |

# 2 Principle of metasurface holography

In this section, we discuss the classification of metasurfaces, the procedure for designing metasurface holograms and three major categories of phase-only holography, amplitude-only holography and combined amplitude/phase holography, by considering the capability of recording the light information on metasurfaces.

## 2.1 Classification of metasurfaces

In recent years, the concept of using metasurfaces for controlling the spatially variant optical properties has enabled a plethora of emerging functions within an ultrathin dimension. One of the pioneering works was carried out by Capasso's group. By judiciously designing an array of V-shape plasmonic nanoantennas to generate an interfacial phase discontinuity that can cover the entire $2\pi$ range, their team was able to control the light trajectory after passing through such metasurfaces [11]. Meanwhile, several excellent reviews on the recent developments of metasurfaces can be found in the literature [10, 42–44].

For a better illustration of the process of combining the holography principle with the metasurfaces as the recording media, we classify below the metasurfaces into four categories based on the constituent materials and operating mechanisms.

(1) Plasmonic metasurfaces: Plasmonic metasurfaces are based on meta-atoms made from metallic nanostructures, whose optical responses are governed by the supported localized plasmon polariton resonances [13]. In the presence of a time-varying external field, the collective motion of the conduction band electrons of the metal can be described as a Lorentz oscillator, which features a resonant peak in the displacement amplitude (polarizability) around the resonance frequency accompanied by a rapid phase shift over the spectral width of the resonance. This resonant phase behavior is the basis of several phase gradient metasurface devices. The most popular strategy uses basic antenna elements (i.e. V-shape [11], Y-shape [45], C-shape [46], etc.) to build up more complex meta-atoms. The meta-atom separation should be smaller than the resonance wavelength to avoid the undesired diffraction effect, but the meta-atoms are still sufficiently separated so that the near-field-mediated interactions are small. Alternatively, based on Babinet's principle, one can also design metasurfaces with complementary structures, i.e. subwavelength apertures in a metallic sheet. The disadvantage of such plasmonic metasurfaces is the low efficiency in transmission, but such a limitation can be conquered by applying metal-insulator-metal schemes working in a reflection configuration, or a more complicated multilayer design [47].

(2) All-dielectric metasurfaces: In contrast to their plasmonic counterparts, dielectric metasurfaces consist of an array of high-index dielectric scattering particles of size comparable to the wavelength of light. The dielectric metasurfaces promise very high efficiencies due to lower absorption losses and easy-tuning scattering properties. Arrays of silicon dimers [48], nanodisks [49] and nanofins with high aspect ratios [50] have been demonstrated for phase, amplitude and polarization control. On the one hand, the dielectric metasurfaces usually exhibit prominent spectral selectivity and high-quality factor (Q-factor) of resonances. On the other hand, some dielectric metasurfaces act as non-resonant phase shifting elements due to their structural birefringence associated with the anisotropic effective refractive index [50]. It has also been proposed that the scattered fields generated by the induced electric and magnetic dipoles in dielectric particles can form the basis of sophisticated antenna arrangements to achieve some interesting scattering effects on demand.

(3) Geometric metasurfaces based on the Pancharatnam-Berry (PB) phase principle: The spatial control of the polarization state of light inevitably introduces nontrivial spatially-varying phase distributions, known as a PB phase. Such a phase is geometric in essence, which is unrelated to the dynamic (propagation) phase that accumulates along the optical propagation path. Specifically, if two parts of a uniformly polarized wavefront are altered to a common polarization state along two different paths on the Poincaré sphere (polarization state space), a relative phase emerges between the two polarization states which is equal to half of the solid angle enclosed by the path [51]. The PB phase represents the evolution of polarization conversion history, so the clockwise and anti-clockwise evolution would flip the sign of such a geometric phase. Such a phase can be obtained by using anisotropic, subwavelength metallic/dielectric scatterers with identical geometric parameters but spatially varying orientations. Nevertheless, in order to achieve a large scattering cross section, the meta-atoms must be resonantly excited. Interestingly, the desired spatial phase tuning mechanism remains totally decoupled from the spectral tuning (e.g. by the size) of the resonators.

(4) Huygens' metasurfaces: The Huygens' principle states that every point on a wavefront can be viewed as a secondary source of wavelets that spread out in the forward direction, thus fulfilling the rigorous boundary condition. In 2013, experimental demonstrations of a kind of Huygens surfaces at microwave frequencies were achieved by satisfying the impedance matching condition with multiple layers of metallic components [52]. To reduce the complexity, the main features of spectrally overlapping, crossed electric and magnetic dipole resonances of equal strength were employed by using high-permittivity all-dielectric scatters [53–55]. The corresponding surface polarizabilities are related to transmission and reflection by the following expression [56]:

$$j\omega\alpha_e^{\text{eff}} = \frac{2(1-T-R)}{(1+T+R)\sqrt{\mu/\varepsilon}}, j\omega\alpha_m^{\text{eff}} = \frac{2\sqrt{\mu/\varepsilon}(1-T+R)}{(1+T-R)}, \quad (1)$$

where $\omega$ represents the angular frequency; $\mu$ and $\varepsilon$ are the permeability and permittivity of the free space and $R$ and $T$ are the complex reflection and transmission coefficients, respectively. By obtaining the required equivalent electric and magnetic polarizabilities for an ideal Huygens' metasurface, full transmission-phase coverage of $2\pi$, low intrinsic losses and almost 100% transmission efficiency can be achieved.

Indeed, metasurfaces are generally hybrid devices, in the sense that they can use the abovementioned mechanisms, such as the PB phase, Mie scattering, Fano resonance, Kerker diffusion and so on, which can add more flexibility to accomplish the smart control of phase, amplitude, polarization and angular momentum.

## 2.2 Procedure for designing metasurface holograms

In general, the procedure for designing and displaying metasurface holograms consists of the following steps, which are also shown in Figure 1 [57]: (i) formulating the mathematical models of the object and of the hologram; (ii) the digital synthesis and numerical calculation of the CGH, resulting in an array of complex numbers that represent amplitudes and phases of interference patterns in the hologram plane; (iii) encoding the phase and/or amplitude information on the physical recording medium (metasurface), in which stage the mathematical holograms are converted into the pattern of building blocks that control the optical properties; (iv) fabricating the metasurfaces through nanofabrication techniques, such as electron beam lithography (EBL), focused ion beam (FIB), nanoimprinting (NI), atomic layer deposition (ALD) and so on and (v) reconstruction of the holographic image by a conventional optical scheme. In CGH, both the image recording and reconstruction procedures are achieved without the need for a reference beam. This is in contrast to the traditional optical holography methods.

For the mathematical modeling in the first step, the objects are approximated as compositions of elementary diffracting elements, such as point scatters, segments of 2D or 3D curves, polygonal mesh or wire-frame models in 3D computer graphics, etc., or described by analytical models. To simplify the description, we assume the monochromatic illumination of objects. The case of multichromatic illumination can be treated as a superposition

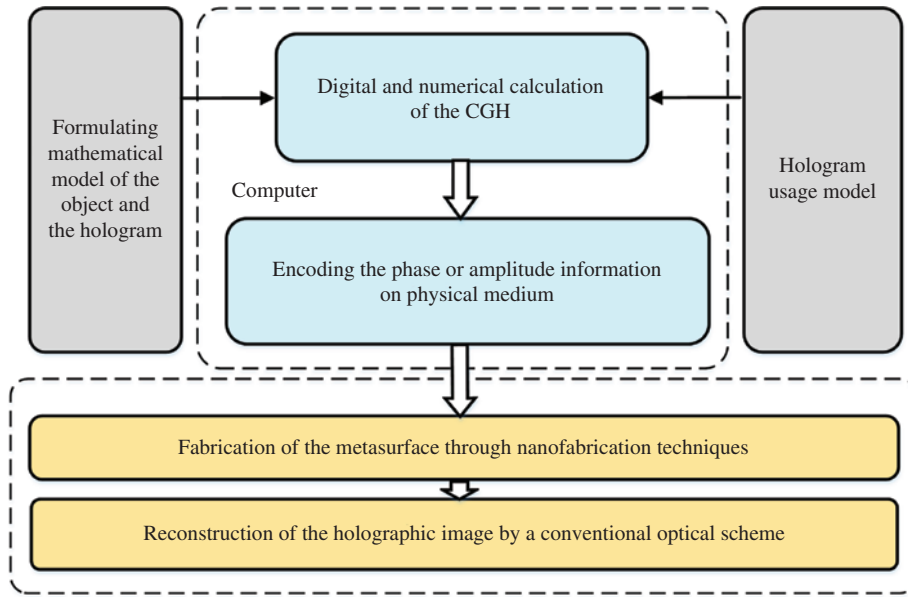

**Figure 1:** Typical procedures required for the design and demonstration of optical metasurface holography.
One important step is the suitable selection of the phase encoding method, which influences the subsequent fabrication step.

of the monochromatic illumination with different wavelengths.

The relation between the scattered wave $\Gamma(\xi, \eta, \zeta)$ over an arbitrary surface defined by its coordinates $(\xi, \eta, \zeta)$ and the object wave $A_{obj}(x, y, z)$ can be described by a wave propagation integral over the object surface or volume $S_{obj}$ [58]

$$\Gamma(\xi, \eta, \zeta) = \iiint_{S_{obj}} A_{obj}(x, y, z) T(x, y, z; \xi, \eta, \zeta) dx dy dz. \quad (2)$$

The wave propagation integral kernel $T(x, y, z; \xi, \eta, \zeta)$ depends on the spatial disposition of the object and the observation surface. The reconstruction of the object can be described by a back-propagation integral over the observation surface $S_{obs}$ given by

$$A'_{obj}(x, y, z) = \iiint_{S_{obs}} \Gamma(\xi, \eta, \zeta) T'(\xi, \eta, \zeta; x, y, z) d\xi d\eta d\zeta, \quad (3)$$

where $A'_{obj}(x, y, z)$ is the reconstructed object wave as seen from the observation surface and $T'(\xi, \eta, \zeta; x, y, z)$ is a kernel reciprocal to $T(x, y, z; \xi, \eta, \zeta)$. Thus, the hologram synthesis requires computation of $\Gamma(\xi, \eta, \zeta)$ through $A_{obj}(x, y, z)$, which is defined by the object description and illumination conditions. Subsequently, the computation results of $\Gamma(\xi, \eta, \zeta)$ are recorded on a physical medium in a form that enables interaction with light for visualizing or reconstructing $A'_{obj}(x, y, z)$ according to the above equation. With different kernel functions, we can achieve Fresnel holograms or Fourier holograms related to different diffraction ranges through discretization. Meanwhile, the sampling interval defines the angular dimensions $(\theta_x, \theta_y)$, of the reconstructed image according to the well-known diffraction relationships [59] given by

$$\theta_x = 2\pi\lambda/\Delta\xi; \quad \theta_y = 2\pi\lambda/\Delta\eta, \quad (4)$$

where $\Delta\xi$ and $\Delta\eta$, respectively represent the lattice constants in the orthogonal directions. The CGH can only be correctly reconstructed by applying the above sampling law. Such criteria can be used for evaluating the performance of holography. More details of the CGH algorithms can be found in [60, 61].

Mostly, it is useful to represent the calculated complex amplitude of calculated hologram in the exponential notation as $A \cdot \exp(i\varphi)$. Metasurface holograms with the capability of recording both the amplitude and phase information of light scattered from the object are desirable for yielding faithful optical images with low noise. According to the development of metasurfaces and various CGH algorithms, the metasurface holograms may be classified into three categories: phase-only holography, amplitude-only holography and combined amplitude/phase holography. A brief classification of the following representative works can be found in Table 1.

## 2.3 Phase-only holograms

In conventional phase-only holograms, the optical thickness of the recording medium is spatially engineered to obtain the required phase of the wave. This can be done either by varying the refractive index of the medium, or the physical thickness, or both. Phase-only media include thermoplastic materials, photoresists, bleached photographic materials, media based on photopolymers, and so on [59]. The first demonstration of holography with metamaterials used multilayer I-shape nanostructures for the local modulation of the refractive index based on effective medium theory [26]. However, such strategy using 3D structured volumes is extremely challenging for nanofabrication and the reconstruction quality of the hologram is less satisfying.

Nowadays, the phase-only metasurface holography is widely employed by using an iterative or point source algorithm in order to optimize the uniformity of the intensity. By adding random phase masks to mimic the diffuse reflection of objects, one can calculate the phase-only CGHs by neglecting the amplitude information. Among the various types of metasurfaces, geometric metasurfaces have shown superior phase control due to the geometric nature of their phase profile related to spatially varying orientations. Huang et al. [57] demonstrated metasurface holograms composed of nanorod arrays based on the PB phase principle with full 3D image reconstruction. For circularly polarized (CP) incident light, the desired local phase shifts are generated in the opposite handedness polarization state of the scattered light, thus achieving the optical reconstruction of the holographic image, as shown in Figure 2A. Due to the subwavelength pixel pitch, the zero-order on-axis 3D reconstruction can potentially be achieved with very high resolution and wide field of view, free of multiple-order diffractions and twin images that are persistent problems in conventional holography. The antenna orientation-controlled PB phase leads to a better broadband performance and helicity switchable property [57]. For this kind of 3D holography, it is easy to verify that the reconstruction distance is inversely proportional to the incident wavelength under the paraxial approximation. By taking advantage of the accurate phase control with PB metasurfaces and the high conversion efficiency of the reflect-arrays that incorporate a ground metal plane into the design, a reflection-type metasurface hologram with 80% efficiency was demonstrated, without the need of a complicated fabrication process (Figure 2B) [27]. Metasurfaces composed of nanoaperture arrays based on Babinet's principle were also used to achieve phase-only holography (Figure 2C) [62]. The advantage of these geometric metasurface holograms is that they are very robust against fabrication errors. Indeed, the hologram is capable of tolerating a fabrication imperfection up to 10% noise, which comprises the shape deformation of the rectangular aperture as well as the phase noise [62].

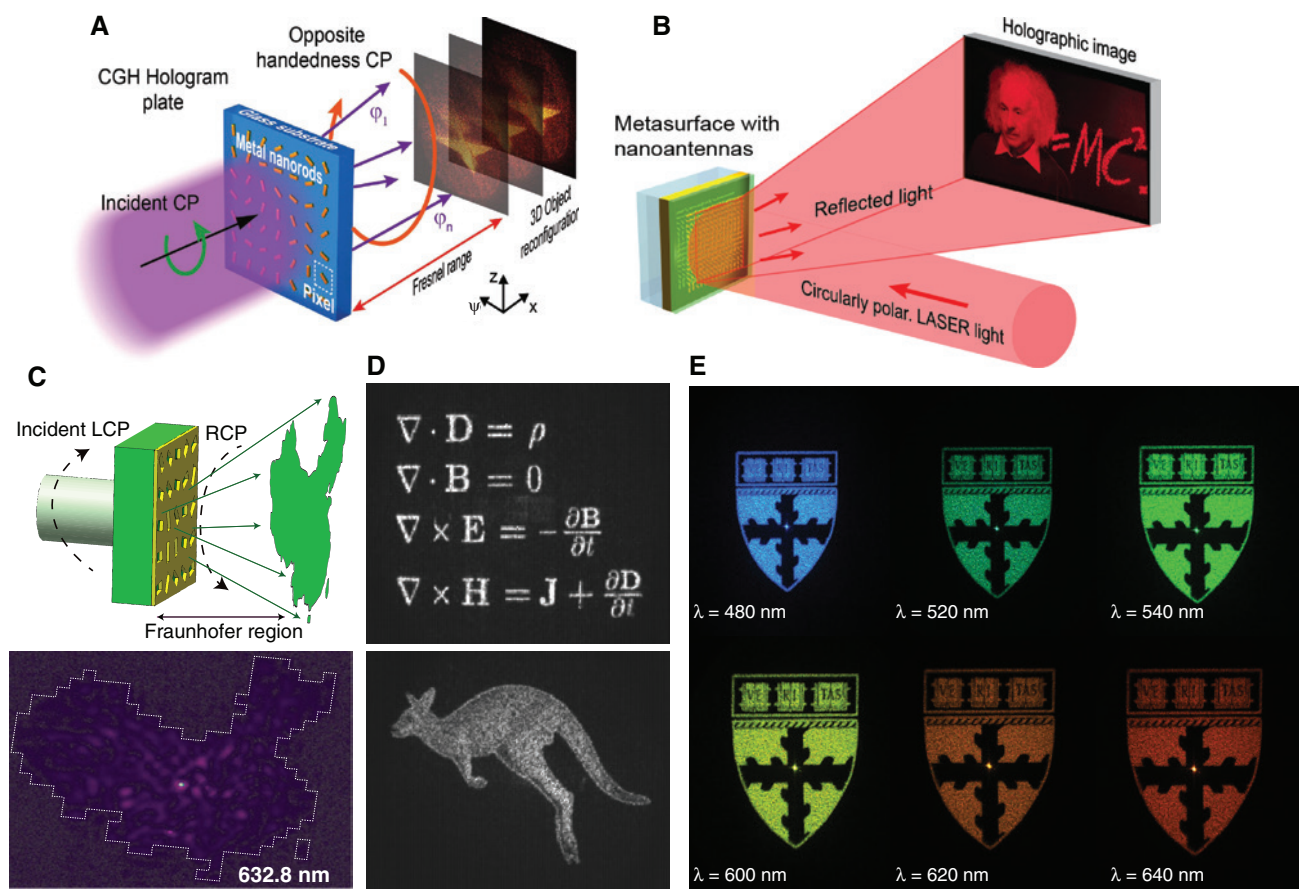

**Figure 2:** Phase-only metasurface holography.
(A) 3D on-axis transmission-type metasurface hologram composed of gold nanorod arrays [57]; (B) 2D off-axis reflection-type metasurface hologram utilizing a metal-insulator-metal layer system for high efficiency [27]; (C) metasurface composed of nano-aperture arrays based on Babinet's principle [62]; (D) dxperimentally obtained images using Huygens' metasurface holograms composed of nanodisks [28]; (E) Reconstructed optical images at different wavelengths covering the entire visible spectrum by using the dielectric metasurface holograms composed of the Si nanofins [63]. Reprint permission obtained from [27, 28, 57, 62, 63].

The all-dielectric geometric metasurface opens another possibility for achieving high efficiency holography in transmission mode. A dielectric metasurface can work efficiently as a half-wave plate for circular polarization conversion by optimizing the geometry parameters with form birefringence, whereas the phase modulation only depends on the orientation of the azimuthal angle of the structure. By using the Si nanopost arrays with spatially-varying orientations, the multicolor holographic images devoid of high-order diffraction and twin-image issues at multiple $z$-planes were achieved in a past work [64]. By leveraging the recently developed Huygens' metasurface, silicon nanodisks with different radii were chosen as the basic meta-atoms to realize phase hologram [56]. In order to ease the fabrication challenge, a seven-level phase map was discretized in equal steps of the nanodisk radii instead of equal steps in phase values. Transmission efficiencies of up to 86% with high fidelity over a wide frequency range were demonstrated. Similarly, the subdiffraction lattices of the silicon nanopillars feature a size-dependent phase delay, which can cover $2\pi$ phase variations. Wang et al. [28] produced grayscale, high-resolution metasurface holograms with transmission efficiencies of over 90% at a wavelength of 1600 nm (Figure 2D). The design approach is also applicable to other materials with high refractive indexes, such as Ge, GaAs, $TiO_2$, diamond, etc. For example, amorphous titanium dioxide ($TiO_2$) was successfully used to extend the working wavelengths spanning the visible spectrum (Figure 2E) [63]. Furthermore, it was reported that meta-atoms designed based on Huygens' principle with a multi-fold rotational symmetry had a polarization-independent response. The distortion induced by the mismatched lattices was also found to be very weak [65].

## 2.4 Amplitude-only holograms

Recently, different schemes for amplitude-only modulation have been proposed. In amplitude-only media, the controlled optical parameter is the local transmission or reflection amplitude. Here, we refer to the amplitude-only media with unit cells assuming only two values in transmission or reflection as "binary media". The simplest strategy for the binary amplitude modulation is based on the process of converting the continuous interference terms of the reference beam and object beam $A_0 A_r^* + c.c.$ into a binary function. Butt et al. [66] achieved binary amplitude holography based on the scattering of carbon nanotubes, as shown in Figure 3A. The amplitude holography information is defined only by their spatial degrees of freedom (0 and 1). However, twin images can appear in the reconstruction. Other groups created binary amplitude holograms based on the optical scattering of the plasmonic nanoparticle [68, 69]. Furthermore, Qiu et al. experimentally demonstrated the accurate manipulation of light by using a random photon sieve to realize a uniform, twin-image free and high diffraction-efficiency hologram, as shown in Figure 3B. By considering the diffraction issues from the subwavelength nanohole, and proceeding with a genetic search algorithm, the performance of the polarization-independent binary amplitude holography was successfully optimized [29]. Similar to the binary amplitude modulation, binary phase modulation with 0 and $\pi$ can also reproduce holographic images with satisfactory quality. Topological insulators, such as $Sb_2Te_3$ thin films, which possess unequal refractive indices in the surface layer and bulk, represent ideal candidates as an intrinsic resonant cavity. The optical path length and phase shifts of an output light beam from the cavity can be enhanced through internal multi-reflections. By printing the binary holograms into the $Sb_2Te_3$ thin films through direct laser writing, Gu et al. [70] demonstrated a 60-nm ultrathin binary hologram.

However, the use of binary holograms is quite inefficient in terms of the medium information storage capacity. However, in amplitude media, in principle, the degrees of freedom related to a transmission (reflection, refraction) amplitude may be used as well. For example, by using a microscopic description of the nanoapertures within the metal film of different sizes to tune the transmission coefficients, Walther et al. [71] achieved multilevel amplitude holography and binary amplitude holography at two wavelengths (Figure 3C) [67]. Such apertures perforated in a metal film can be approximated as an assembly of dipolar emitters.

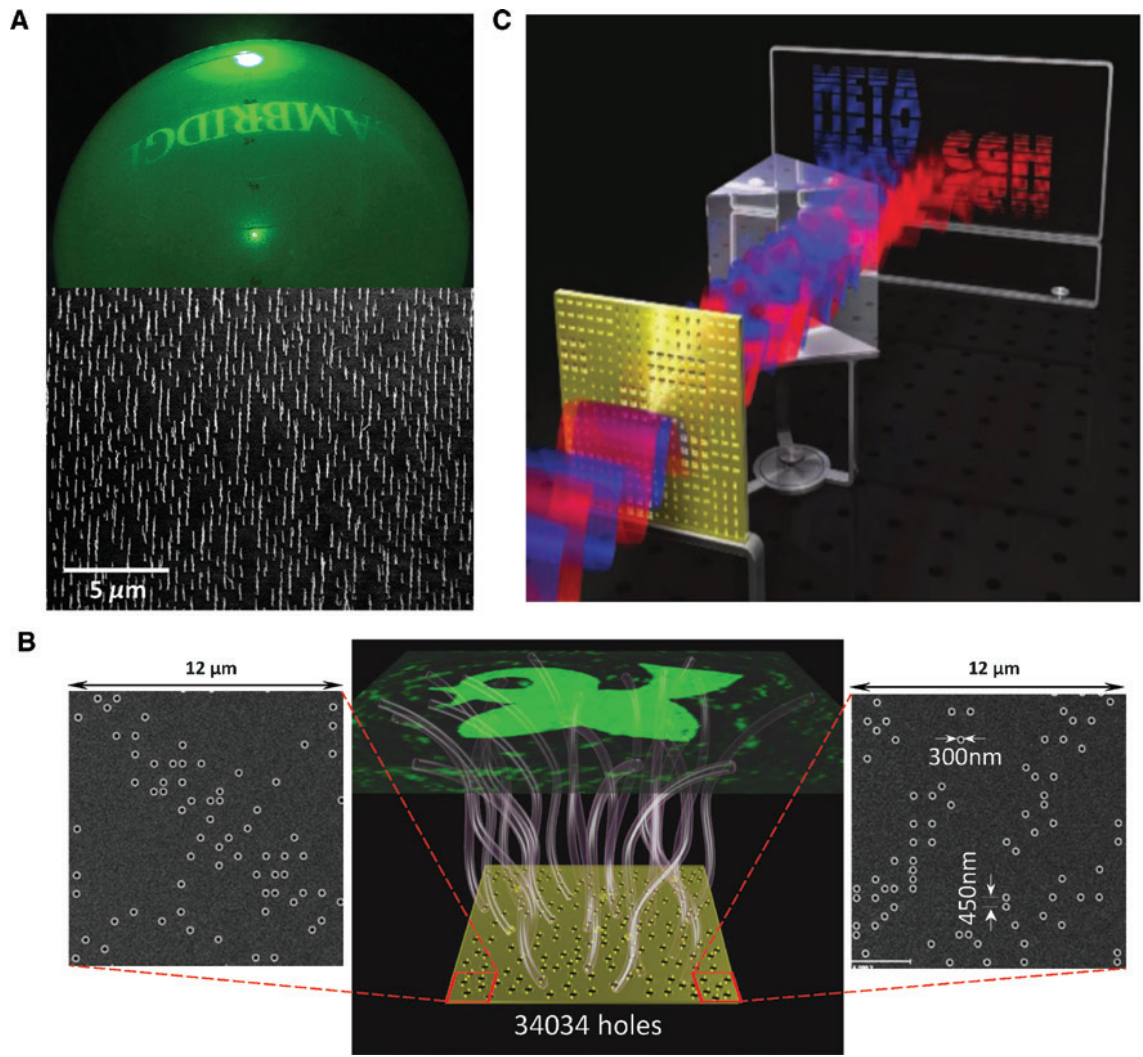

**Figure 3:** Amplitude-only metasurface holography.
(A) Reconstructed binary amplitude hologram (top) by using carbon nanotubes (bottom) [66]; (B) binary amplitude hologram by using a randomly patterned photon sieve [29]; (C) schematic of a metasurface hologram with the nanoaperture arrays perforated in a metal film with binary amplitude modulations at two wavelengths [67]. Reprint permission obtained from [29, 66, 67].

## 2.5 Complex amplitude holograms

An on-demand metasurface capable of imprinting arbitrarily complex wavefronts on an incident light can, in principle, realize any holographic images. A complex device with the desired functionalities can be achieved by simultaneously modulating the phase and amplitude information. In this case, the hologram can reconstruct the object without losing any information. Further, various reflection properties of the object surfaces, such as specular reflection and Phong shading, can be designed with optimized performance [72]. The complex amplitude approach has unique advantages, including higher precision and reduced noise.

Shalaev et al. [30] achieved metasurface holography with delicately designed V-shape nanoapertures based on Babinet's principle, which requires an extra lookup table for controlling the local phase and amplitude. For a linearly polarized incident beam, the complementary design effectively blocks the direct transmission of the co-polarized component as the rest of the sample remains opaque due to the continuous metallic film, as shown in Figure 4A. For such plasmonic resonance tuning metasurfaces, the spectral bandwidth is quite limited; it also introduces extra phase noise into the holographic image when illuminating at different wavelengths [30]. By adopting the C-shape split-ring resonators (CSRRs) as the basic unit, the amplitude and phase of the outgoing orthogonally-polarized linear wave can be simultaneously manipulated by varying the geometrical parameters (radius $r$, split angle $\alpha$ and orientation angle $\theta$). It has been demonstrated that the amplitude of the outgoing $y$-polarized component follows a simple $|\sin(2\theta)|$ dependence with respect to the orientation angle $\theta$, whereas the phase shift remains invariant [46]. This facile way of controlling the amplitude without interfering with the phase profile greatly simplifies the complex amplitude design. A novel complex amplitude modulation hologram with simultaneous five-level amplitude modulation and eight-level phase modulation has been achieved accordingly, as shown in Figure 4B [31]. In contrast, an independent phase and amplitude modulation scheme can be achieved by using a geometric metasurface, where the phase can be tuned by the azimuthal angle and the amplitude can be separately modulated by tuning the shape.

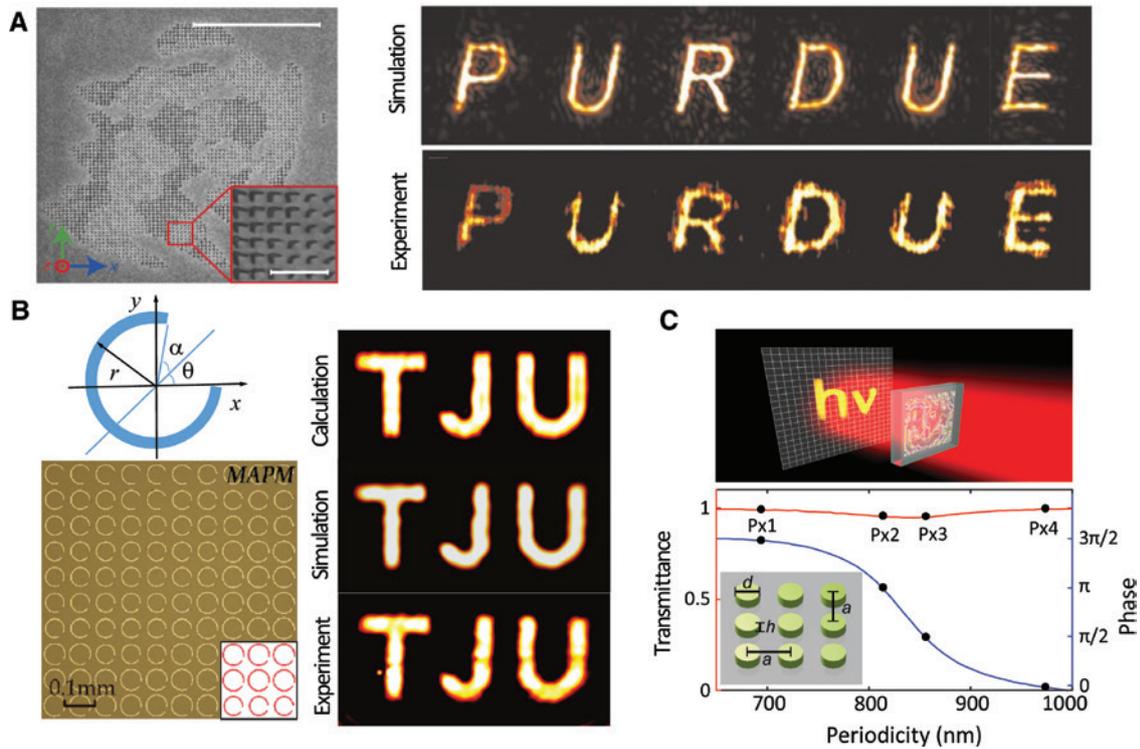

**Figure 4:** Complex amplitude metasurface holography.
(A) SEM image of a metasurface made of V-shape nanoapertures with two-level amplitude modulation and eight-level phase modulation (left); simulation and experimental results of the reconstructed image (right) [30]. (B) Optical image of a metasurface made of C-shape antennas with simultaneous amplitude and phase modulation (left), together with theoretical calculation, numerical simulation and experimental results of the reconstructed holographic image (right) [31]. (C) Schematic image of a Huygens' dielectric metasurface for the complex amplitude holography and its transmission coefficients by modulating the lattice periodicities [32]. Reprint permission obtained from [30–32].

Another strategy for realizing the complex amplitude modulation of metasurface holography is by applying the concept of Huygens' metasurfaces. Given that a variation of the lattice periodicity would lead to a spectral shift of the resonance (while keeping the diameter and height to be constant), the spatial transmittance phases introduced at a particular frequency would then be dependent on the range of the selected lattice periodicity. However, the amplitude modulation is quite small and the phase range cannot cover the total $2\pi$ range (Figure 4C) [32].

## 3 Holographic multiplexing

The possibility of optimizing the enormous space-bandwidth product and information capability of metasurface holograms provides a strong motivation for the development of suitable multiplexing techniques. Traditionally, most of these techniques were developed for volume holographic recording systems with Bragg-based selectivity [73]. In contrast, metasurfaces may circumvent some of the limitations of earlier techniques employing photorefractive crystals, which may suffer from the photobleaching effect in the reading processing, and establish the feasibility of multiplexing without introducing distortion in the reconstruction [74].

In the following, we discuss three prospects of metasurface holograms with potential applications: color holography, polarization multiplexing and hybrid multiplexing algorithms. Such multiplexing technologies can be used as a platform for low-cost, high-performance and large-capacity holographic displays and data storage systems.

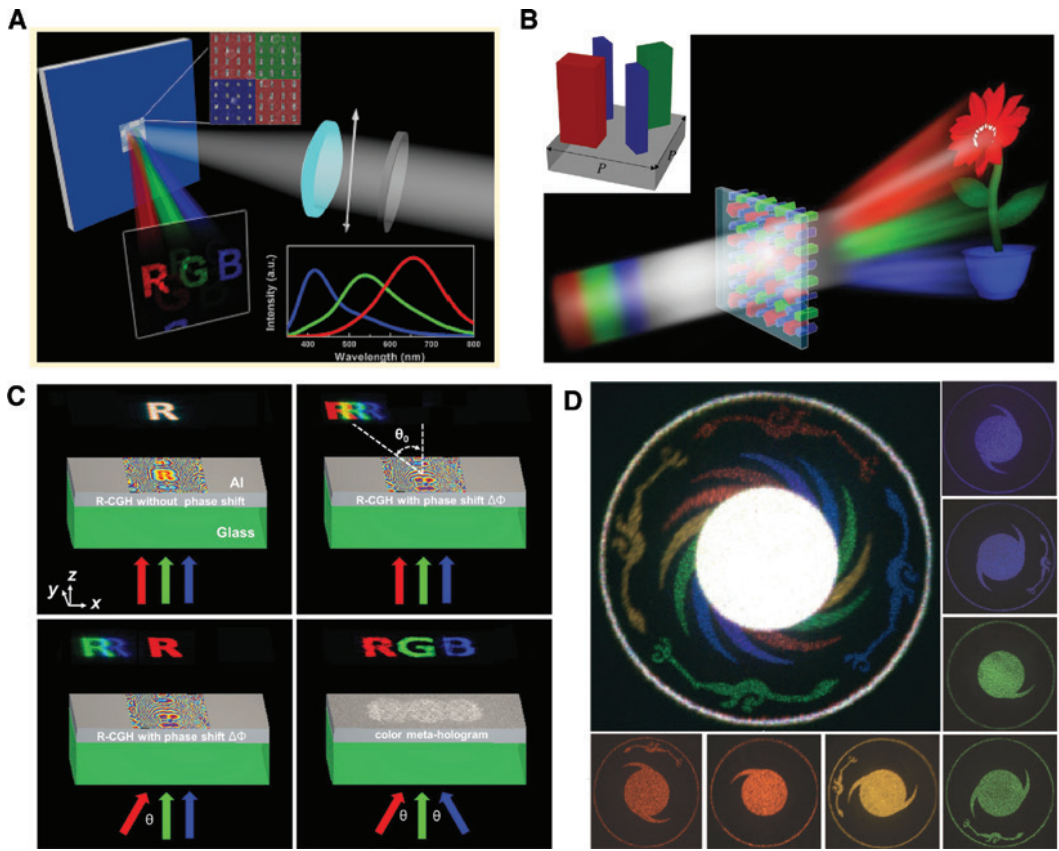

**Figure 5:** Metasurface color holography.
(A) Schematic illustration of the multicolor plasmonic metasurface hologram composed of the aluminum nanorods under the linearly polarized illumination. One meta-atom consists of four subpixels [75]. (B) Schematic illustration of multicolor dielectric metasurface hologram consisting of multiplexed Si nanoblocks. The hologram independently projects distinct color images: a red "flower", a green "peduncle" and a blue "pot" [33]. (C) Principle and experimental realization of a color metasurface hologram composed of the nanoaperture arrays with the tilted off-axis strategy [76]. (D) Seven-color holographic image for Sun Phoenix, which is reconstructed from a metasurface composed of the nanoaperture arrays [77]. Reprint permission obtained from [33, 75–77].

## 3.1 Color holography

Multicolor holography is one of the most pursued applications for metasurfaces, which is very important for display and wavelength division multiplexing. The major challenge in realizing color holography is to independently record and reconstruct the amplitude and phase information on different color components. In this case, extra design freedoms or complex encoding methods in each unit cell must be considered.

The most intuitive approaches for color holography with metasurfaces are based on spatial multiplexing schemes. For example, Tsai et al. demonstrated a metasurface hologram considering a two-level phase modulation for each primary color. Each pixel is made of four subpixels: one for blue, one for green and two for red to compensate for the lower reflectance in the red spectral range, as shown in Figure 5A. Within each subpixel, all aluminum nanorods have the same dimensions, but the subpixels from different pixels can have different dimensions to yield either 0 or π phase shifts for a given color. Taking into account the wavelength dependence of the diffraction angle, different images can be reconstructed at specific locations with a predefined size. With this technique, they obtained resonances of the aluminum rods within a narrow bandwidth, thus facilitating the implementation of the multicolor scheme [75]. However, with this strategy, several nanoantennas need to be grouped to obtain an effective response within the pixel, which degrades the image resolution and reduces the information density as well as the viewing angle of the hologram [75, 78]. Usually, an $M$-level phase control at $N$ discrete wavelengths requires $M \times N$ types of resonators. When $M$ increases, the dimension differences of the resonators decrease, thereby increasing the fabrication difficulty, especially for shorter wavelengths [33]. Nevertheless, the resonator types can be reduced to $N$ by applying the PB phase principle. The dielectric metasurfaces made of three kinds of Si nanoblocks within a subwavelength unit-cell were demonstrated for transmission-type, high-efficiency color holography. These metasurfaces are capable of simultaneous wavefront manipulation of the three primary light colors, as shown in Figure 5B [33]. Full phase control is easily achieved by changing the in-plane orientations of the corresponding nanoblocks to serve as narrow-band half-wave plates at the non-overlapping discrete wavelengths for the circularly polarized illumination. Achromatic and highly dispersive meta-holograms were fabricated to demonstrate the wavefront manipulation with high resolution [33].

Another approach uses geometric metasurfaces consisting of subwavelength nanoslits with spatially varying orientations in thin metal films via the utilization of the broadband effect of the PB phase in combination with the off-axis CGH reconstruction scheme. Each RGB component of the color object can be considered as many self-illuminating point sources at the corresponding wavelength. Based on the Huygens-Fresnel principle, the CGH of each color component (R-CGH, B-CGH and G-CGH) can be individually calculated by superimposing the optical wavefronts from all the point sources. Additional phase shifts are encoded into the CGH of each color component. By introducing the corresponding tilted incident angle illumination, the desired color images can be reconstructed, as shown in Figure 5C. Due to the dispersion, only the desired image with the right color can be correctly projected above the hologram. The scattered light beams at the designed angles are then superimposed to form the final multicolor image. Nevertheless, all the other "unwanted" images appear blurred in other positions due to the broadband effect. Another factor influencing the quality of the reconstructed image is the color dispersion with different magnifications of the reconstructed images for the three primary color components. Such an effect can be eliminated in advance through the numerical zero-padding technique [76]. Based on the same principle, Luo et al. demonstrated a seven-color "Sun Phoenix" metasurface holographic image, as shown in Figure 5D. Especially, the larger color gamut describes a broader spectrum, in which the hologram can be reproduced in the color space by using seven lasers of different wavelengths. The reconstructed image showed good quality and a wide range of viewing angles, but with low efficiency [77].

Despite the above demonstrations, two of the challenges for multicolor operation are efficiency and crosstalk. Narrow resonances are generally favorable for minimizing the crosstalk between different color images, but this also implies that multiple components should be used, resulting in degraded viewing angles. While the color holography based on broadband resonances usually requires delicate tuning with off-axis illumination, recent demonstrations showed the ability of metasurfaces to overcome chromatic aberration [79, 80], which can be potentially helpful in the development of compact metasurface color holographic devices.

## 3.2 Polarization multiplexing

Another technique for multiplexing, which uses the polarization states of the light, cannot be easily achieved in traditional holography due to the lack of polarization-sensitive natural materials. Past studies used birefringent metasurfaces composed of two superimposed independent transverse nanoantenna arrays arranged in orthogonal directions to achieve polarization multiplexing [68, 69, 81]. As shown in Figure 6A, there is very little interference or cross-talk in the far-field radiation [69]. Using the polarization as an additional degree of freedom, it is possible to design multi-wavelengths and polarization-sensitive metasurface holograms. Using this technique, two independent holograms were sampled with square grids of nanorods and paired nanospheres for two colors at the orthogonal polarization states [68]. The local plasmonic resonances along the cross-axes were addressed by orthogonal linear polarization states. As the nanorod arrays showed both longitudinal and transversal resonance peaks in the red and blue regions, they produced the same images at two different colors, whereas the nanosphere arrays produced a single image at one color [68]. The helicity switchable property of the PB phase was applied for the spin multiplexing scheme. By using two spatially multiplexed reflection arrays of silver nanoantennas, the resulting real and virtual holographic images with the spin dependence of incident photons can be reconstructed and switched, as shown in Figure 6B [82].

With the freedom of designing both the cross-section and rotation angles of dielectric resonators, metasurface platforms that provide complete control of polarization and phase with subwavelength spatial resolution were experimentally achieved by using elliptical or rectangular nanoposts [34, 83]. The general relation between the electric field of the input and output waves at each pixel can be expressed using the Jones matrix. Indeed, any desired symmetric and unitary Jones matrix can be realized using a birefringent metasurface if the polarization dependent phase shift ($\phi_x$, $\phi_y$) and the azimuthal angle $\theta$ can be chosen freely. The freedom provided by the proposed platform to simultaneously control the polarization and phase of light allows for the implementation of the polarization-switchable phase hologram with two distinct patterns for x- and y-polarized light, as shown in Figure 6C [34]. Capasso et al. [83] further extended the method to allow the superposition of two independent and arbitrary phase profiles on any pair of orthogonal states of polarization-linear, circular or elliptical-relying only on such birefringent metasurfaces. Crucially, the desired phases (combining both propagation and geometric phases) can be imparted on any set of orthogonal polarization states by simultaneously modifying the cross-section and angular orientation of each element. To demonstrate this arbitrary phase control for different polarizations, a metasurface that can encode separate holographic images for RCP and LCP light was fabricated. The near-field phase profiles yielding the far-field intensity images of a cartoon cat and dog were computed using iterative phase retrieval, as shown in Figure 6D. In contrast to geometric phase metasurfaces, only sections of the far-field can contain independent images for each spin [82]. While using the polarization multiplexing method presented here, the phase profiles imparted on each circular polarization and, consequently, the resulting far-fields, can be completely decoupled [83].

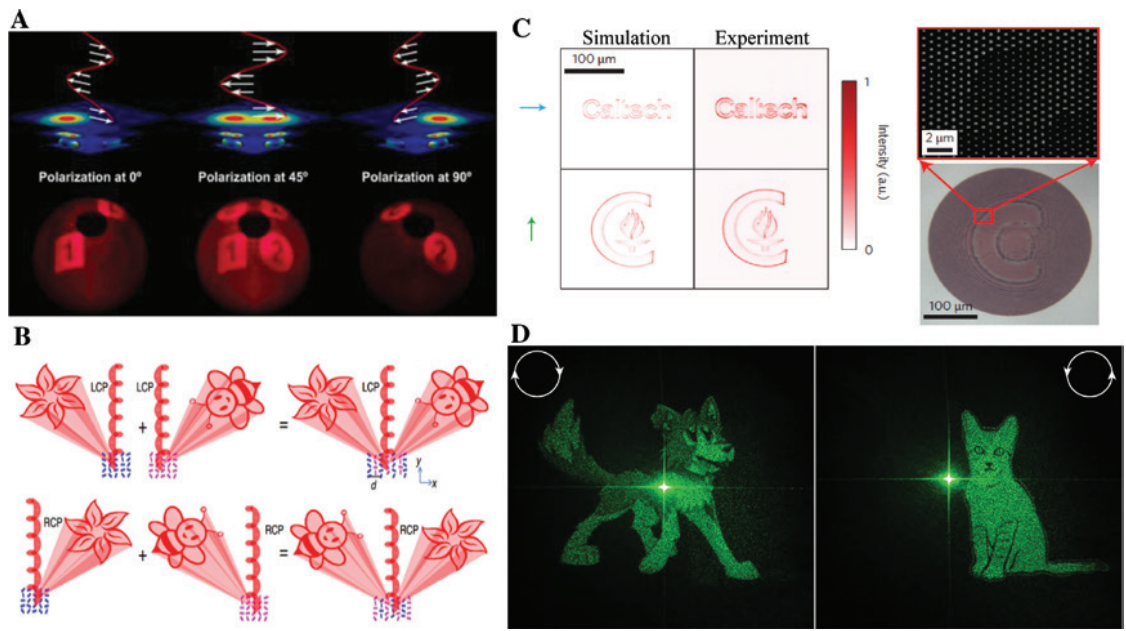

**Figure 6:** Polarization multiplexing.
(A) Linear polarization multiplexing by superimposing the two orthogonal arrays of the nanorods within one metasurface [69]. (B) Circular polarization multiplexing by interleaving two sets of the nanorod arrays based on the PB phase principle [82]. (C) Simultaneous polarization and phase modulation for demonstrating two independent holographic images by using the tailored elliptical posts for orthogonal linear polarizations [34]. (D) Experimental realization of a cartoon dog and cat with tailored Si nanofins for the orthogonal circular polarization multiplexing [83]. Reprint permission obtained from [34, 69, 82, 83].

### 3.3 Algorithm

The two categories above discussed the strategies of holographic multiplexing from the perspective of the physical mechanisms. In the following, we introduce some recently developed holographic algorithms for multiplexing. In general, the design of the metasurface holographic multiplexing is mathematically cast into a combinatorial optimization problem, with different display channels, such as polarizations, positions, angles and wavelengths.

The concept of the detour phase served as the core concept for the earliest CGHs, where the amplitude and the phase of the optical field were imposed by an array of apertures on the screen. The wavelets diffracted along a given direction from two neighboring apertures with distance $D$ were phase-shifted relative to each other by the amount $\Delta\varphi = 2\pi D/\lambda \sin\theta$, whereas the dimension of each aperture determines the amount of light passing through it [84, 85]. Capasso et al. designed a metasurface consisting of dielectric ridge waveguide groups (DRWs), made of amorphous silicon, on a glass substrate. The interaction of the DRWs with the incident light is highly polarization-dependent because of the deep-subwavelength width and asymmetric cross-section. A diffraction condition where most of the transmitted light was funneled into the first orders (±1) was achieved by adjusting the DRW design parameters (width, height, separation and lateral dimension) of the meta-element. The desired phase profile was then converted into a spatial distribution of displacements $D(x_m, y_m)$, which was equivalent to a phase-only hologram. Furthermore, by using nanofins to form the supercells based on the geometric phase principle, the required phase map was achieved through the displacement of each supercell, and chiral holographic multiplexing was demonstrated [86].

For achieving multiple recording channels, an effective way to handle various images is to combine geometric metasurfaces with a simplified synthetic spectrum holographic algorithm. Such synthetic spectrum is composed of individual hologram spectra with different linear phase shifts. The selection of suitable parameters of the superimposed objects is crucial to the success of high-quality holographic multiplexing and reduced cross-talk between different images. Multiple hybrid multiplexing schemes were demonstrated, with position, polarization and angle channels. Thereby, each image can be reconstructed with a unique key [35]. Meanwhile, the reconstruction images can be reversed or rotated through a coordinate transformation. Another method for multiplane holographic multiplexing was proposed by using the 3D Fienup algorithm that constructed an iterative loop between multiple target object planes and the hologram plane with an optimization to obtain the required single-phase profile [87]. Similar to the well-known GS algorithm, the phase profile of the hologram was optimized with an amplitude replacement in each object plane and an amplitude normalization applied in hologram plane [88]. Such a 3D Fienup algorithm was characterized by introducing a feedback function used for the amplitude replacement at the object planes which can be expressed as $T_n = T + |T - T'_n|\kappa$. In that equation, $T$ is the amplitude of the target object, $T'_n$ is the amplitude calculated from the $n$th iteration loop and $T_n$ is the result of the feedback function, which is used to replace $T'_n$. By appropriately choosing the feedback operation, the convergence speed can be increased significantly. As a result, one can get an optimized phase-only hologram that allows the reconstruction of different target objects in different planes within the Fresnel range. Hence, the multiplexing capacity $P_{max}$ can be significantly enlarged. Note that the estimation of $P_{max}$ depends on the similarities of the recorded images, the overlap of the angular spectra and the proper choice of key parameters. Alternatively, by employing an orthogonal technique with base functions satisfy the Dirac-delta-relationship in the superposition process, the expected image can be exactly recovered free of cross-talk [35].

# 4 Surface wave holography

In the previous section, we discussed holograms for which the reconstruction is formed by free-space optical waves. However, surface waves, such as surface plasmon polaritons (SPPs), play an important role in several fields of science and technology. The subwavelength confinement of SPPs has been revolutionizing the methods by which to control light at the nanoscale. For the surface holography, there is a fundamental difference because the SPP wavefronts are reconfigured or coupled from far field radiations through the interaction of metasurfaces. This opens up interesting new possibilities as it provides novel and powerful methods for controlling the near-field distribution and connection of SPPs. In the following, we introduce the principle of surface wave holography, together with various applications based on such techniques.

## 4.1 Principle

For mimicking the traditional holography principle, interference between a plasmonic reference and a plasmonic/free-space object beam was suggested. In 2011, Ozaki et al. [89] achieved SPP color holography by recording the interference patterns of the object waves and reference waves with white light illumination from high index medium onto a surface. The SPP wave was then converted by such metasurface hologram into a radiative free-space light field, which reconstructed the object wave. Given that such a technique provides a natural separation between the reference wave and object wave, surface wave holography is advantageous in terms of background-free reconstruction.

Another method for surface holography defines equal phase grooves $\varphi_{SPP}(x, y) = \varphi_{obj}(x, y) + 2m\pi$ that can be etched in the metal surface (Figure 7A) [90, 92]. This assumes the existence of fictitious object beam sources for the recording process, which are reconstructed by the plasmonic reference wave. However, such a method does not address the coupling issue of the interacting SPPs [90]. Similarly, a four-fold multiplexed hologram was designed by incorporating the SPP propagation with polarized scattering, as shown in Figure 7B [36]. The reference SPP beam propagating towards the metasurface hologram from different directions is scattered by the corrugations to generate a free-space wavefront with predefined amplitude and phase distribution [36]. By approximating the nanoapertures as an ensemble of electric dipoles distributed on the surface, continuous phase modulation of SPPs was achieved, thus fulfilling the condition for surface wave holography (Figure 7C)

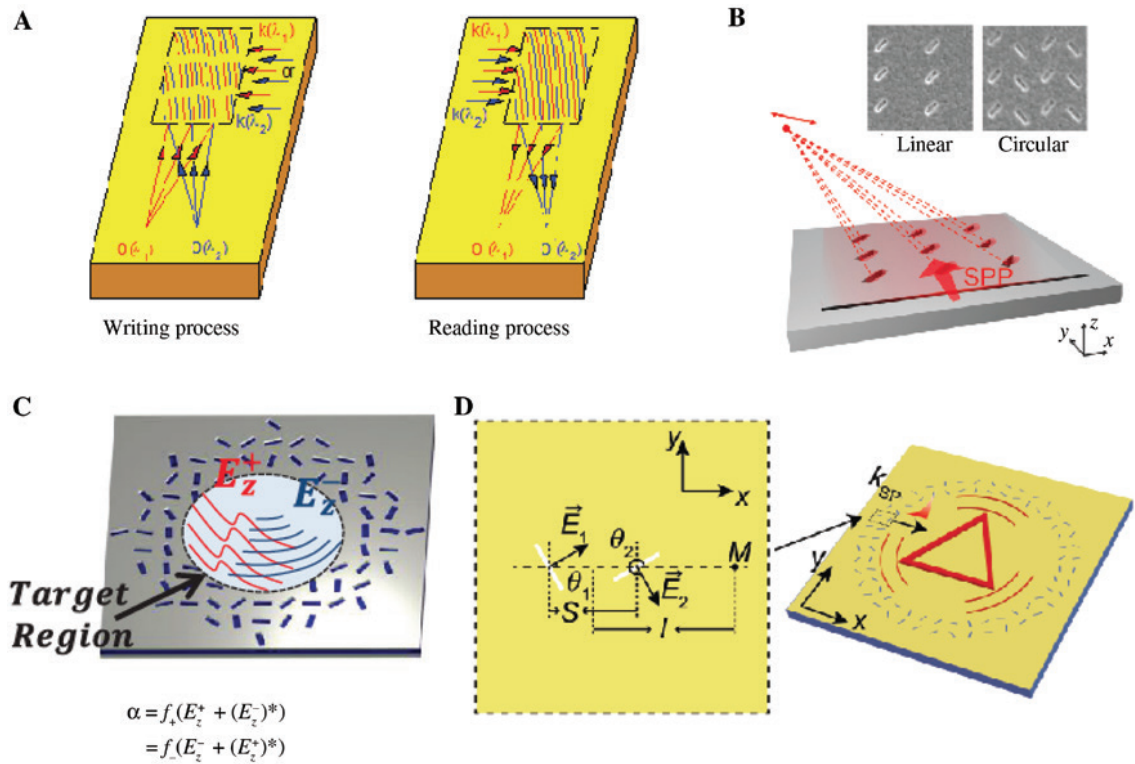

**Figure 7:** Surface wave holography.
(A) Schematic of the writing and reading process of in-plane surface holography for multiple wavelengths [90]. (B) Design of the metasurface for the four-fold polarization controlled holographic multiplexing by coupling the SPPs to far-field radiation [36]. (C) Matching rule of surface holography by using the metasurfaces composed of nanoapertures to gain the independent spin-selective control of SPPs [37]. (D) Schematic views of the metasurface unit composed of a pair of nanoslits, for the simultaneous modulation of the amplitude and phase of SPPs for surface wave holography [91]. Reprint permission obtained from [36, 37, 90, 91].

[37]. This is made possible by combining the holographic principle with a geometric phase matching scheme, which determines the orientation profile of the nanoapertures from the superimposition of the target SPP profiles for the two spins. The geometric phase, $\pm 2\alpha$, at each particle is matched to $\arg(\partial_x \pm i\partial_y)E_z$ instead of simply $\arg(E_z)$, by considering the vectorial nature of the propagation wave if a precise control of profiles is needed. Furthermore, the spin-enabled control of orbitals is potentially useful for constructing dynamic motion pictures with a series of picture frames. The maximum intensity can be tuned to occur at a polarization-selective angle $\xi$, by multiplying $e^{\pm i\xi}$ on the LCP and RCP target profiles [37]. Meanwhile, it is desirable to introduce the SPP amplitude control simultaneously to increase the imaging quality. A control strategy to directly generate complex SPP holographic profiles by using slit-pair resonators was proposed (Figure 7D) [91]. By patterning the orientation angles ($\theta_1$, $\theta_2$) of such slit-pair resonators while keeping the distance $S$ conserved, the phase of the excited SPPs can be freely controlled by the summation of angles ($\theta_1 + \theta_2$) with the sign determined by the incident spin, while the amplitude is proportional to $\sin(\theta_1 - \theta_2)$. Similarly, the interference could be gradually tuned from constructive to destructive by designing the target profiles with a different initial phase [91]. Such tunability is particularly attractive in designing the SPP force to manipulate the motion of nanoparticles, which can be trapped or released by the intensity tunable closed-loop SPP profiles.

Four issues need to be addressed in order to realize surface holography. First, coupling an SPP wave from a free-space wave requires momentum matching between the two wave vectors [93]. Second, the SPPs have a vectorial nature; thus, their excitation conditions and their amplitude and phase should be defined accordingly. Third, in the near-field, all wavefronts or diffraction orders interfere together and the finite propagation distance should be considered. Finally, the holographic scheme means that it ideally works only for a single wavelength. However, they can tolerate a wavelength shift of about ±5% with acceptable performance for most general schemes that do not provide dedicated dispersion compensation [94].

## 4.2 Applications of surface wave holograms

Recent advances in the spatial and spectral shaping of SPPs are based on the realization of plasmonic holograms [25, 94]. We like to emphasize that such a holographic approach is very different from transformation optics. For transformation optics, the parameters are judicially chosen to create 2D distributions of the wavevector and the Poynting vector, as well as to ensure impedance matching between the adjacent unit cells [95]. A holographic approach can relax the strict conditions of extreme permittivity/permeability parameters as it only requires tailoring the phase and amplitude of the generated SPPs [94]. For this vision to bear fruits, new plasmonic metasurface holograms are needed so that these can act as optical interconnects between nanoscale plasmonic volumes and free-space optical elements.

Such holographic couplers can be utilized to achieve the selective detection of the orbital angular momentum of light. The metasurface hologram was designed with the interference patterns of an object free-space vortex beam with a certain topological charge and a reference plasmonic beam. When free-space beams were shined on such hologram, which controlled the spatial shape of SPP beam, by using this concept, it was possible to detect the original orbital angular momentum carried by the free-space beam [96]. Other strategies utilized interference between a reference plasmonic beam and an object free-space beam to reconstruct a desired free-space beam wavefront in the far-field [25]. Such holographic couplers have potential applications in various areas of integrated optics by combining metasurfaces with a standard commercial photodiode. Even though the inherent loss of the SPP wave is fundamental and acts as a major inhibitor for a variety of plasmonic applications, the ability to generate arbitrary SPP profiles in a controllable manner based on a holographic principle is a crucial step in designing plasmonic imaging and lithography devices. We note that, apart from fully controlling the SPP wavefront, metasurface holography can also be utilized for tailoring the energy spectrum of the SPP excitation [94]. This may open up exciting opportunities for the temporal shaping of SPP waves.

## 5 Nonlinear holography

The abilities of metasurfaces to promote light-matter interaction and manipulate local optical polarizations are ideally suited to tailor nonlinear optical effects [97]. In addition to the success of metasurfaces in the linear optical regime, the concept of phase tailoring metasurfaces was recently extended to the nonlinear regime, thus enabling both coherent generation and manipulation of new frequencies.

Inspired by the concept of the linear PB phase elements, a nonlinear plasmonic metasurface was demonstrated, which allowed a continuous control of the phase of the local effective nonlinear polarizability [98]. It was shown that the nonlinear polarizability of the meta-atom can be expressed as $\alpha^{n\omega}_{\theta,\sigma,\sigma} \propto e^{(n-1)i\theta\sigma}$ and $\alpha^{n\omega}_{\theta,-\sigma,\sigma} \propto e^{(n+1)i\theta\sigma}$, respectively, for the $n$th harmonic generation with the same or opposite spin ($\pm\sigma$) compared with that of the fundamental wave. The relative phase factors $(n-1)\theta\sigma$ and $(n+1)\theta\sigma$ of the nonlinear waves depend only on the orientation angle $\theta$ of the meta-atom. Thus, two different continuous phases were imposed on the harmonic generation signals of opposite spins. Such geometric nonlinear PB phases enabled a continuous control over the phase change from 0 to $2\pi$. According to the selection rules for harmonic generation of circular polarized fundamental waves, the meta-atom allows only harmonic orders of $n=lm\pm1$, where $m$ represents the $m$-fold rotational symmetry and $l$ is an arbitrary integer. Although the orientation angle controls the nonlinear phase for certain harmonic generation orders, it does not affect the linear optical response for rotational symmetries with $m>2$. Hence, the linear and nonlinear optical properties of the rotationally symmetric antennas can be decoupled from each other [98].

Combining the concept of nonlinear metasurfaces with mature holography optimization techniques, it is possible to convert one fundamental wave into multiple beam profiles or images using a single metasurface. Ye et al. selected split ring resonators (SRRs) as meta-atoms due to their strong polarization properties in the linear regime as well as the high second-harmonic generation efficiency. The SRR possessed one-fold mirror symmetry, which can be used for investigating second harmonic generation (SHG) due to the lack of centrosymmetry. For a fundamental beam of spin $\sigma$ illuminating upon such a metasurface, the transmitted fundamental beam of $-\sigma$ acquires a linear PB phase of $2\sigma\theta$, whereas the SHG signals with spin $\sigma$ and $-\sigma$ acquire a nonlinear PB phase of $\sigma\theta$ and $3\sigma\theta$, respectively. Thus, by patterning orientation angles $\theta(\mathbf{r})$ of each individual SRR, the phase profiles can be arbitrarily manipulated to optimize the construction of three independent image intensity profiles, as shown in Figure 8A [38]. Note that the three different images are independently read out by selecting the frequency and spin of the transmitted light. In this way, multi-channel information capacity can be obtained for a single metasurface by utilizing nonlinear optical processes.

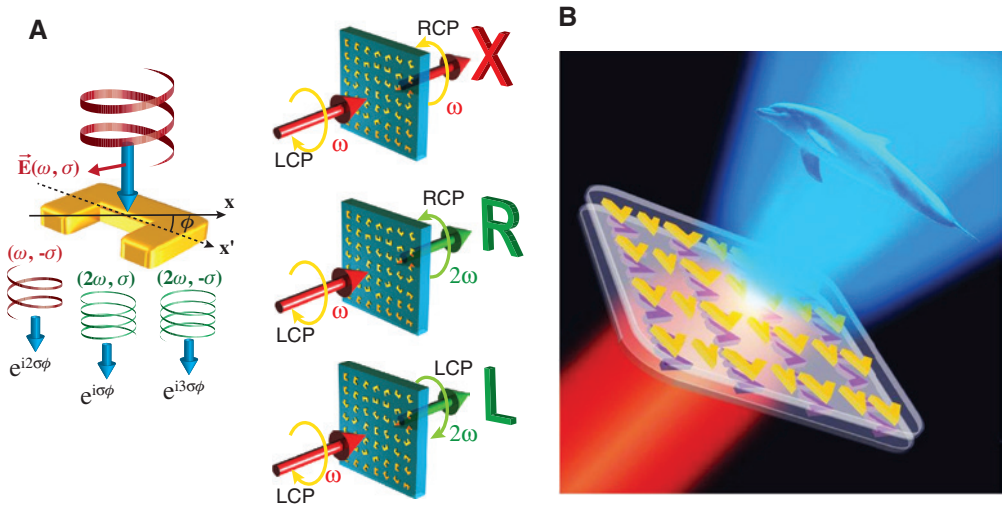

**Figure 8:** Nonlinear metasurface holography.
(A) Schematic of the nonlinear phase modulation principle using the split-ring resonators with C1 symmetry, together with the spin- and wavelength-dependent holographic images at different channels [38]. (B) Nonlinear metasurface hologram composed of double-layer V-shaped antennas for the third-harmonic generations of holographic images [39]. Reprint permission obtained from [38, 39].

Another strategy uses multilayer metasurfaces composed of V-shape nanoantennas for nonlinear holography (Figure 8B) [39]. By approximating the nonlinear nanoantennas as point dipoles with effective third-order nonlinear susceptibility $\chi_{eff}^{(3)}$, the third-order material polarization (oscillating at $3\omega$) is given by [39]

$$\mathbf{P}^{(3)}(3\omega) \propto \chi_{eff}^{(3)}(3\omega,\omega)[a(\omega)\mathbf{E}_1 e^{i\varphi(\omega)}]^3, \quad (5)$$

where $a(\omega)$ is the near-field enhancement, $\varphi(\omega)$ is the phase-shift of the fundamental beam and $\mathbf{E}_1$ is the incoming wave. Therefore, the third-harmonic field $\mathbf{E}_3$ will acquire phase shift of $3\varphi(\omega)$, whereas an additional relative phase shift may be added depending on the resonant nature of $\chi_{eff}^{(3)}$. This approach was used to embed two active nanoantenna layers with different holograms in each layer. Note that each image was recreated only by the properly polarized input beam and emanates from a single layer of nanoantennas within the double layer metasurfaces. As an example, the authors demonstrated the reconstruction of the Hebrew letters Aleph and Shin for vertical and horizontal polarizations, respectively [39].

Actually, in a general theoretical framework employing the effective nonlinear susceptibility, the polarization vector $P_i$ containing both linear and nonlinear terms can be written as

$$P_i = \varepsilon_0 \sum_j \chi_{ij}^{(1)} E_j + \varepsilon_0 \sum_{jk} \chi_{ijk}^{(2)} E_j E_k + \varepsilon_0 \sum_{jkl} \chi_{ijkl}^{(3)} E_j E_k E_l + \dots. \quad (6)$$

From there, it is clear that every combination of polarizations of the impinging and outgoing beam acquires a specific phase factor that depends on the nonlinear susceptibilities and phase shifting of each element, thus allowing the full control of $0$–$2\pi$ phase range [99]. Such kind of nonlinear metasurfaces may provide more possibilities for controlling higher-order nonlinear holography.

Nonlinear metasurface holography is a promising candidate for applications in anti-counterfeiting, hidden security identification features, multidimensional optical data storage and optical encryption. Indeed, by using symmetry-controlled high-harmonic generation from metasurfaces, it is now possible to engineer both the polarization and wavefronts of high-order harmonic generation efficiently.

## 6 Active metasurface holography

Active metasurfaces have taken off as an emerging subfield of metasurfaces, stimulating a gradual transition of metasurface holography from passive to active elements. Such devices play a crucial role in various nanophotonic systems [9]. Metasurfaces integrated with various functional materials, such as phase change materials [43] (germanium antimony tellurium alloy $Ge_2Sb_2Te_5$, vanadium dioxide $VO_2$) and 2D materials [100] (graphene, borophene), can extend their exotic passive properties by allowing pixel-level independent control of active components. Through various modulation methods, such as thermal excitation, voltage bias, magnetic field, optical pump or mechanical deformation, the local optical properties can be judiciously tuned to adapt to the operational conditions, which may ultimately empower dynamic capabilities for computational imaging, wireless communication and so on.

Many efforts have been done by incorporating novel functional materials into metasurfaces. Graphene oxide (GO) has emerged as an active material once it is reduced back towards graphene or reduced graphene-oxide (rGO), yielding contrast in optical and electronic properties. A thermally digitalized photoreduction of GOs in photopolymers by femtosecond laser enables the control of the refractive index of rGOs. Through an area-by-area digitalization to achieve the refractive index modulation, Li et al. [40] demonstrated the multilevel phase holography with a diffraction efficiency of 15.88%, as shown in Figure 9A. Meanwhile, the reversible reduction and oxidation of GOs by switching the polarity of electrical stimulus might be used for updatable holograms. In addition, the spectrally flat refractive-index modulation of the photoreduction process enables its application for wavelength-multiplexed full-color 3D images [40]. Another promising method uses a thin film of chalcogenide phase-change material encapsulated by the dielectric layers for the hologram panel and excimer laser lithography. The representative phase-change materials, such as vanadium dioxide ($VO_2$) or germanium antimony tellurium alloy ($Ge_2Sb_2Te_5$, GST), have been thoroughly investigated for integrated active optical devices. The phase change of GST films is non-volatile, repetitive and easily controlled by both the electric and optical stimuli. According to the material properties of the GST between amorphous or crystalline phase, significant changes in the refractive index and extinction coefficients can be reached. With that, the resonance condition after the insertion of the GST film into a dielectric medium can be shifted from the original resonance condition, allowing a large phase shift of the reflection coefficient (Figure 9B) [41].

By applying the isotropic strain to metasurface holograms on a stretchable polydimethylsiloxane (PDMS) substrate, the hologram image can be reconfigured [101]. The Huygens-Fresnel Principle has shown that for a given stretch ratio $s$, the electric field alters to $E'(x',y',z') = E'(sx, sy, s^2z) = E(x,y,z)e^{ik(s^2-1)z}$. Therefore, if a metasurface hologram is stretched by a factor of $s$, the hologram image is enlarged accordingly, and the hologram image plane moves away from its unstretched $z$ position as $(S^2-1)z$. By leveraging the fact that the image plane position changes when the substrate is stretched, the reconfigurable metasurface holograms can be obtained. If the reconstructed hologram image is observed at a fixed plane, it can be changed upon stretching the metasurface, resulting in a different reconstructed image (Figure 9C) [101].

Additionally, active elements (e.g. varactors and diodes) were utilized on metasurfaces to demonstrate the dynamic EM wave controls at microwave frequencies. For example, by incorporating an electric diode into the unit cell of the metasurface, the scattering state of each individual unit cell can be controlled. Cui et al. demonstrated a programmable 1-bit coding metasurface hologram that can generate arbitrary holographic images by programming the input control voltages via the field programmable gate array (FPGA). The state of each unit cell of the coding metasurface can be switched between "1" and "0" by electrically controlling the loaded diodes. Consequently, the binary amplitude holograms can be reprogrammed through digitalization. The reconfiguration time is completely determined by the FPGA clock rate [102]. Such reprogrammable metasurface holograms can be readily extended to exhibit multiple bits and simultaneous phase and amplitude modulations that can lead to versatile devices with adaptive and rewritable functionalities. All the functionalities of metasurface holography mentioned above are briefly summarized in Table 2.

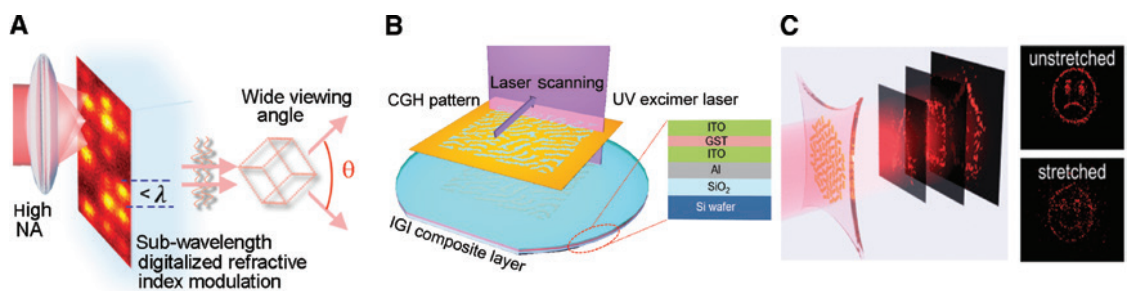

**Figure 9:** Active metasurface holography.
(A) Active metasurface hologram composed of the functional graphene oxide materials for achieving 3D holography with wide viewing angles, which is rewritable by pumping with femtosecond laser pulses [40]. (B) Active metasurface hologram integrated with the phase change material GST, which can be switched between the crystallization or amorphous state [41]. (C) Mechanically stretchable metasurface hologram for switchable image reconstructions at a fixed plane [101]. Reprint permission obtained from [40, 41, 101].

# 7 Holographic-related techniques and perspectives

The metasurface holography demonstrated so far possesses high design flexibility owing to its diverse modulation mechanisms and structural building blocks. The holography principle is used as a tool to solve an inverse engineering problem to excite either free-space beams or surface waves with any desirable field distribution. Nevertheless, several challenges remain, in particular in terms of theoretical modelling, metasurface design and real-time reconfigurability. In addition, advanced nano/fabrication techniques are highly desirable for realizing subwavelength features over large areas to obtain high-quality metasurface holograms.

(1) Theoretical modeling

In general, the existing holographic modeling methods (point source algorithm, polygonal mesh algorithm, GS algorithm, genetic algorithm, etc.) can be directly applied to metasurfaces [59]. The new degrees of freedom provided by the metasurfaces can conquer traditional problems faced by holographic related techniques. Despite the new opportunities, metasurfaces continue to face new challenges, as partially discussed in Section 3.3. Nevertheless, to exploit the full potential of the local optical properties modulated by the metasurfaces (e.g. simultaneous complex amplitude and polarization modulation, angular momentum multiplexing, active/switchable modulation and so on), smart algorithms/strategies should be developed for different application occasions. Meanwhile, there is a need for additional scalar/vector theoretical tools (method of moments for antenna analysis, dipolar modeling and so on) that can directly connect shape and material properties of meta-atoms to the complex macroscopic values for more accurate calculations, especially by considering the evanescent waves [103]. Furthermore, accurate, high precision and fast calculations of CGHs are highly on-demand.

(2) Metasurface design

For the sake of future development, it is desirable to incorporate metasurfaces with other platforms for achieving novel physical mechanism. Studies on materials with low losses, durability and CMOS compatibility for metasurfaces have become very prominent in recent years, and several material systems were used to demonstrate all-optical, non-volatile, switchable and high-resolution holography. By designing scattering elements to provide the degrees of freedom to simultaneously engineer the local amplitude, phase and polarization response on an ultrathin interface, it might be possible to provide a suitable platform to realize all types of vector holograms. Additionally, as metasurfaces are frequently illuminated using normal incidence, the collective modes based on in-plane electric dipole momentums are typically involved. However, the collective excitations involving the out-of-plane electric dipole moments have received much less attention. A natural approach to probe such modes is to use optical illumination away from normal incidence upon nanopillars/multilayer metasurfaces. Alternatively, a 3D full-wave analysis is required for such complex system. To achieve a complete metasurface design theory, it is desirable to investigate such out-of-plane dipole moments.

(3) Real-time reconfigurability

Despite the rapid progress in active metasurfaces, the simultaneous realization of real-time reconfigurability, high efficiency and large FOV for the scattered light has yet to be achieved. We remark that the reconfigurable metasurface hologram may be realized by using the thermal effect phased arrays [24], electro-optical modulations [23], liquid crystals [104] and programmable gate arrays [102]. Several well-established switchable diodes may facilitate the frequency scaling from microwave to other spectral regions, for instance, the Schottky diode at the terahertz and the thermal $VO_2$ diode at infrared and visible frequencies. To extend the dynamic metasurface holography to higher frequencies through dynamically modifying the phase change materials, there is a need to further develop multi-beam control techniques by laser stimulus to enhance the rewritable/real-time control rate. Additionally, incorporating cutting-edge nanofabrication techniques for developing pixelated nanoelectrode and nanogenerator/controller might be a key step in enabling future intelligent devices with reconfigurable and programmable functionalities, such as novel camera modules for cell phones and laptops, wearable displays for augmented reality, metrology, 3D imaging and lasers for various innovative portable/handheld instruments.

(4) Applications

The universality and simplicity of the holographic strategies can offer promising opportunities for various practical applications. Here, we only name a few of the important related techniques.

   (a) Optical information processing. Metasurface-based CGHs can be used as spatial filters, beam-shaping elements, deflectors, beamsplitters and optical interconnects and can also be used for optical communications, high-performance computing (multiplications) and for information display.

   (b) Optical metrology. Holographic interferometry is a technique that enables the measurements of static and dynamic displacements of objects with optically rough surfaces at an optical interferometric precision [105]. The metasurface holograms can be used to detect optical path length variations in transparent media, which allows, for example, a fluid flow to be visualized and analyzed. They can also be used to generate contours representing the form of the surface or the isodose regions in the radiation dosimetry. Such a technique can be widely used to measure the stress, strain and vibration in engineering structures.

   (c) Chemical/biological sensors. The metasurface hologram made from a modified material can be utilized to interact with certain molecules to generate a change in the fringe periodicity or refractive index. Hence, the color of the holographic reflection can be changed [106]. Meanwhile, many of the existing chemical and biological sensors' principle can also be applied to metasurfaces.

   (d) Optical security and counterfeit. The potential of holograms as security features for valuable documents and products has been realized since the early days of holography [59]. The metasurface holograms are ideal candidates as an encrypted pattern because of the convenient and robust processing of data handling and encryption and due to their high-precision and easy usage of multiple control parameters, including polarization, wavelength, positions, materials properties and so on, which can increase the security level.

## 8 Summary

In this overview article, we reviewed the history and development of metasurface holography – from the early stages of holographic image generation to the most recent advanced holographic multiplexing. We presented several approaches that have been used to enable the control of the free space and surface plasmon polariton wavefront based on metasurface holography principles. Here, we only discussed the electromagnetic metasurfaces, while metasurfaces can also be incorporated to achieve the acoustic holography for various sound field applications. However, the acoustic counterpart to the optical holography is beyond the scope of this review.

Metasurfaces based on holographic related techniques are not only expanding our understanding of the underlying physics but also can potentially lead to the discovery of unanticipated phenomena and applications. We anticipate that the advancements in this field are crucial in the realization of any arbitrary beam control and can further stimulate the development of new products by the nanotechnology industry.

**Acknowledgments:** This work is financially supported by the European Research Council Consolidator Grant (NONLINMAT) and the National Key R&D Program of China (No. 2017YFB1002900). L. H. acknowledges the support from the National Natural Science Foundation of China (Grant No. 61775019) and Beijing Nova Program (Grant No. Z171100001117047).

**Additional information**
**Competing financial interests:** The authors declare no competing financial interests.